\documentclass[aps,pre,reprint,onecolumn,notitlepage,groupedaddress]{revtex4-1}
\usepackage{epsfig,graphics,amssymb,amsmath,color,mathtools,subeqnarray,bm,mathtools,upgreek}
\usepackage[colorlinks=true,linkcolor=blue]{hyperref}
\usepackage[sfdefault=cmbr]{isomath}

\def \t{\tensorsym}
\def \lb{\left}
\def \rb{\right}

\def \d{\,\text{d}}

\def \bchi{\boldsymbol{\chi}}

\def \bDelta{\mathbf{\Delta}}

\def \bnabla{\boldsymbol{\nabla}}

\def \bOmega{\mathbf{\Omega}}
\def \bOmegah{\hat{\mathbf{\Omega}}}

\def \bsigma{\boldsymbol{\sigma}}
\def \bSigma{\boldsymbol{\Sigma}}
\def \bsigmah{\hat{\boldsymbol{\sigma}}}

\def \bTheta{\mathbf{\Theta}}

\def \bzero{\mathbf{0}}
\def \tzero{\mathsf{\t 0}}

\def \bA{\mathbf{A}}

\def \bB{\mathbf{B}}

\def \fB{\mathcal{B}}

\def \bC{\mathbf{C}}

\def \bE{\mathbf{E}}

\def \tE{\mathsf{\t E}}

\def \bF{\mathbf{F}}

\def \fF{\mathcal{F}}

\def \tF{\mathsf{\t F}}

\def \bG{\mathbf{G}}

\def \bI{\mathbf{I}}
\def \bbI{\mathbb{I}}

\def \bL{\mathbf{L}}

\def \bM{\mathbf{M}}
\def \fM{\mathcal{M}}
\def \tM{\mathsf{\t M}}

\def \bR{\mathbf{R}}

\def \fR{\mathcal{R}}

\def \tR{\mathsf{\t R}}

\def \bS{\mathbf{S}}
\def \bSt{\tilde{\mathbf{S}}}

\def \tS{\mathsf{\t S}}

\def \bT{\mathbf{T}}

\def \tT{\mathsf{\t T}}

\def \bU{\mathbf{U}}

\def \bUh{\hat{\mathbf{U}}}

\def \fU{\mathcal{U}}

\def \tU{\mathsf{\t U}}
\def \tUh{\mathsf{\t{\hat{U}}}}

\def \bX{\mathbf{X}}

\def \ba{\mathbf{a}}

\def \bb{\mathbf{b}}

\def \bc{\mathbf{c}}

\def \be{\mathbf{e}}

\def \bf{\mathbf{f}}

\def \bn{\mathbf{n}}

\def \bp{\mathbf{p}}

\def \br{\mathbf{r}}

\def \bu{\mathbf{u}}
\def \buh{\hat{\bu}}

\def \but{\tilde{\bu}}

\def \bx{\mathbf{x}}

\def \by{\mathbf{y}}

\begin{document}

\title{Active Stokesian Dynamics}
\author{Gwynn J. Elfring}\email{Electronic mail: gelfring@mech.ubc.ca}
\affiliation{
Department of Mechanical Engineering, Institute of Applied Mathematics,
University of British Columbia
Vancouver, BC, Canada}
\author{John F. Brady}\email{Electronic mail: jfbrady@caltech.edu}
\affiliation{
Division of Chemistry and Chemical Engineering, California Institute of Technology
Pasadena, California, USA}
\begin{abstract}
Since its development, Stokesian Dynamics has been a leading approach for the dynamic simulation of suspensions of particles at arbitrary concentrations with full hydrodynamic interactions. Although originally developed for the simulation of passive particle suspensions, the Stokesian Dynamics framework is equally well suited to the analysis and dynamic simulation of suspensions of active particles, as we elucidate here. We show how the reciprocal theorem can be used to formulate the exact dynamics for a suspension of arbitrary active particles and then show how the Stokesian Dynamics method provides a rigorous way to approximate and compute the dynamics of dense active suspensions where many-body hydrodynamic interactions are important.
\end{abstract}
\maketitle

\section{Introduction}
Active matter is a term used to describe matter that is composed of a large number of self-propelled active ‘particles’ that individually convert stored or ambient energy into systematic motion \citep{schweitzer07, morozov17}. The interaction of many of these individual active particles can lead to complex collective dynamics \citep{ramaswamy10}. Natural examples include a flock of birds, a school of fish, or a suspension of bacteria \citep{toner05}, but active matter may also be composed of synthetic active particles \citep{bechinger16}. These out-of-equilibrium systems are most often in fluids, and so understanding their dynamics and rheology involves a connection between fluid-body interactions and non-equilibrium statistical physics \citep{marchetti13, saintillan18}.

The study of active matter at small scales is complicated by the fact that the Stokes equations, which govern momentum conservation of Newtonian fluids when inertia is negligible, feature a long-range decay of fluid disturbances \citep{happel65}. Because of this, active particles interact through the fluid over distances long relative to their individual size and to properly capture the effect of the fluid in these systems one may need to sum hydrodynamic interactions between \textit{all} bodies, particularly at higher particle concentrations. 

The difficulty of accurately capturing many-body hydrodynamic interactions is well known from the study of suspensions of passive particles, where early efforts to sum hydrodynamic interactions in infinite suspensions were plagued by problems of divergent sums (see for example the long literature on sedimenting particles \citep{davis85}) but eventually overcome by the pioneering work of \citet{batchelor72}, \citet{jeffrey74}, \citet{hinch77}, \citet{obrien79} and others. The Stokesian Dynamics method developed soon after facilitated the efficient dynamic simulation of passive particle suspensions at arbitrary concentrations \citep{brady88}. The essential basis of the Stokesian Dynamics method is a mixed asymptotic approach wherein hydrodynamic forces on particles due to interactions are computed distinctly when the particles are in close proximity versus widely separated. When the particles are widely separated the method sums many-body hydrodynamic reflections between particles through inversion of a truncated grand mobility tensor, whereas when the particles are in close proximity pair-wise additive lubrication forces are used \citep{durlofsky87}. When the suspension is infinite or periodic, a modification of the method introduced by \citet{obrien79} is used to obtain absolutely convergent expressions for the hydrodynamic interactions among all particles, suitable for the numerical simulation of a wide range of problems from sedimentation to rheology \citep{brady88a}. Since its inception the Stokesian Dynamics method has served as a foundational tool for the development of our understanding of suspension mechanics in the last several decades.

Unlike passive suspensions, in active suspensions each active particle in the fluid is endowed with non-trivial boundary conditions due to activity and constantly injects energy into the fluid. Many advances have been made in understanding the dynamics of individual swimming microorganisms (biological and synthetic), from the pioneering work of \citet{taylor51}, through to several detailed reviews of microscale locomotion research \citep{lighthill76, brennen77, lauga09b}. However, in a similar fashion to the early development of the passive suspension literature, the majority of research on collective locomotion of many bodies and active suspensions has emphasized dilute suspensions where swimmer-swimmer interactions are greatly simplified and far-field approximations are still valid \citep{saintillan18}. Interesting phenomena, such as particle clustering (motility-induced phase separation) has been observed for dense suspensions of active particles \citep{bechinger16}, but very often numerical simulation of these suspensions is done with active Brownian particle (ABP) models that neglect hydrodynamic interactions entirely \citep{cates15}. Others have used approaches for active suspensions that only approximate the Stokes equations, such as multiparticle collision dynamics \citep{zottl14} or lattice Boltzmann methods \citep{stenhammar17} that still may not be accurate for very dense concentrations. Results for simplified swimmers in concentrated suspensions display qualitative differences \citep{ishikawa08, evans11, alarcon13, zottl14, matas-navaro14, thutupalli18}. Some argue that hydrodynamic interactions act to suppress phase separation in active matter \citep{matas-navaro14}, while others have shown that hydrodynamic interactions with boundaries can control phase separation \citep{thutupalli18}. A complete understanding of the connection between individual particle activity, the hydrodynamic interactions between many particles that arise as a consequence of this activity, and the role this plays in the macroscopic dynamics of concentrated active suspensions has not been developed. 

As we discuss in the following, the Stokesian Dynamics methodology is easily adapted for the dynamic simulation of suspensions of active particles at any concentration and, as with passive suspensions, is particularly well suited for dense concentrations and periodic boundary conditions. The mathematical structure of the dynamical equations remains essentially unchanged between passive and active particles and so any implementation of the Stokesian Dynamics method for passive particles, may be simply and easily modified for use with active particles. Moreover, we believe this mathematical structure provides an ideal formalism for theoretical analysis of hydrodynamic interactions in active matter much as it has for passive suspensions \citep{brady93a,brady93b}. 

The Stokesian Dynamics method was first adapted for use with self-propelled active particles by \citet{mehandia08}. In their work, they introduce spheres each with a prescribed virtual propulsive force, that interact through a prescribed stresslet whose magnitude sets the size of the virtual propulsive force, and an induced stresslet caused by particle rigidity in a bulk flow. The dynamics of these active spheres was then solved numerically using the Stokesian Dynamics framework. The authors found that near-field interactions appeared important even at low concentrations, as particles tended to cluster, and they found qualitative differences in the dynamics between low and high volume fractions. Despite the novelty, the authors did not specify how the propulsive force arises from the surface boundary conditions nor how to generalize this approach. Shortly afterwards, \citet{ishikawa08} adapted the Stokesian Dynamics framework for use with spherical particles with a prescribed tangential slip velocity, so-called squirmer particles \citep{ishikawa08}. Using their own previous results for two-body hydrodynamic interaction between squirmer particles \citep{ishikawa06}, \citet{ishikawa08} were able to incorporate both near-field interactions and many-body far-field interactions for the study of dense suspensions of (2-mode) squirmer particles. This framework was then used to study the rheology \citep{ishikawa07}, diffusion \citep{ishikawa07b} and coherent structures \citep{ishikawa08b} of these active suspensions. The Stokesian Dynamics framework was then extended for use with passive and active spherical particles with a fairly general surface velocity field (but were individually immotile) that could be linked together to form complex swimming assemblies \citep{swan11}. That machinery was then used to simulate a number of model swimming microorganisms, from pusher and puller swimmers to helical flagella, by using assemblies of spherical particles \citep{swan11}. Recently, the far-field (but not the near-field) approach taken by Stokesian Dynamics, namely constructing mobility tensors by a moment expansion of the boundary integral equations, was extended to arbitrary squirmers \citep{singh15}. Here we show that this previous literature may all be encapsulated by a fairly general theory for the hydrodynamic interaction of arbitrary active particles that can then be solved efficiently with the Stokesian Dynamics approach, particularly for spherical particles.

We begin by developing a general kinematic description of an arbitrary active particle in section \ref{sec:kinematics}. We then show how the reciprocal theorem can be used to yield the exact dynamics for a suspension of $N$ arbitrary active particles in section \ref{sec:dynamics}. We then show how the Stokesian Dynamics technique is used for the approximation and dynamic simulation of these exact equations in section \ref{sec:stokesiandynamics}.

\section{Kinematics of an active particle}\label{sec:kinematics}
\begin{figure}
\center
\includegraphics[width = 0.6\textwidth]{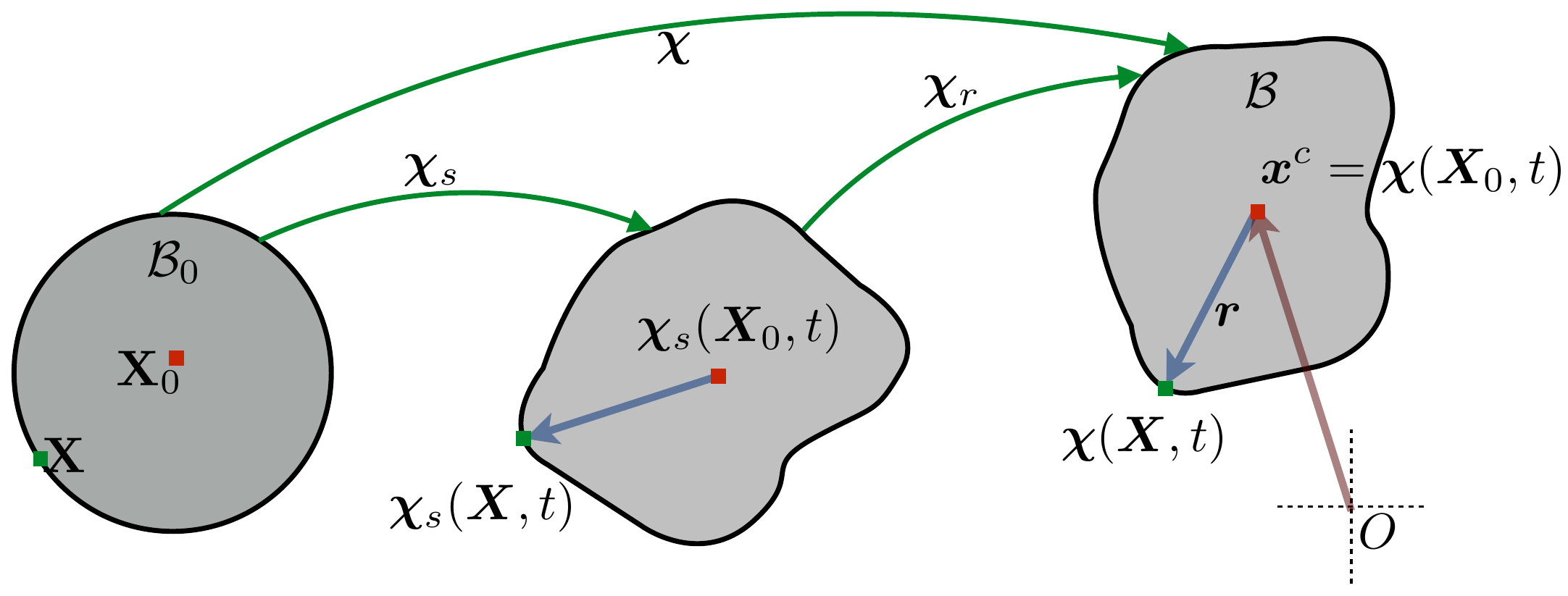}
\caption{Schematic of a deforming active particle.}
\label{particle}
\end{figure}

Consider an active particle identified with the region $\fB$ as shown in figure \ref{particle}. 
Changes in the spatial configuration of the active particle can be described by a map, $\bchi$, from a reference configuration $\fB_0$, such that $\bx=\bchi(\bX,t)$ for $\bx\in\fB$ and $\bX\in\fB_0$. The motion of the body can be decomposed into shape change, $\bchi_s$, which represents the swimming gait of the active particle, and rigid-body motion, $\bchi_r$, which arises as a consequence of interaction with the fluid, so that  
\begin{align}
\bchi(\bX,t) = \bx^c(t) + \bTheta(t)\cdot(\bchi_s(\bX,t)-\bchi_s(\bX_0,t)),
\end{align}
where $\bx^c$ is the translation and $\bTheta$ the rotation (about $\bchi_s(\bX_0,t)$) of the body under the action of $\bchi_r$. Upon differentiation, we obtain the velocity of the body
\begin{align}
\bu(\bx\in \fB) = \bU+\bOmega\times\br+\bu^s,
\label{bcparticle}
\end{align}
where the translational velocity $\bU = d\bx^c/dt$, while the rotational velocity $\bOmega$ is defined by $d\bTheta/dt = \bOmega\times\bTheta$ and $\br = \bTheta(t)\cdot(\bchi_s(\bX,t)-\bchi_s(\bX_0,t))$. The deformation velocity due to shape change is
\begin{align}
\bu^s=\bTheta\cdot\frac{d(\bchi_s(\bX,t)-\bchi_s(\bX_0,t))}{dt}.
\end{align}
where the last term is the deformation velocity in the unoriented configuration,
\begin{align}
\frac{d(\bchi_s(\bX,t)-\bchi_s(\bX_0,t))}{dt}=\bTheta^{-1}\cdot\bu^s\equiv\but^s.
\end{align}
In a purely kinematic description of the activity of the particle we would consider the shape change $\bchi_s$, or simply the surface velocity (in the reference orientation) $\but^s(\bX \in \partial \fB_0)$, to be prescribed and then solve for the rigid-body translation and rotation of the active particle such that momentum (of the particle and of the fluid) is conserved.

\section{Dynamics of active particles}\label{sec:dynamics}
Consider a suspension of $N$ particles, each labeled $\fB_i$ where $i\in[1,N]$, immersed in an arbitrary background flow denoted $\bu^\infty$. The disturbance velocity field generated by the particles is
\begin{align}
\bu' = \bu-\bu^\infty.
\end{align}

Neglecting the inertia of the active particles and of the Newtonian fluid in which they are immersed, the rigid-body dynamics of active particles is governed by an instantaneous force balance
\begin{align}
\tF+\tF_{ext}=\tzero,
\label{forces}
\end{align}
where $\tF$ and $\tF_{ext}$ are respectively $6N-$dimensional vectors of hydrodynamic and external (or interparticle) forces/torques on all $N$ particles.

In general, the hydrodynamic forces may be easily shown, by the reciprocal theorem of low Reynolds number hydrodynamics, to be weighted integrals of the boundary traction during rigid-body motion,
\begin{align}
\tF &= \sum_i\int_{\partial\fB_i}\bu'\cdot(\bn\cdot\tT_{\tU})\d S,
\label{hydrodynamicforce}
\end{align}
where $\bn$ is the normal to the surface, $\partial \fB_i$, pointing into the fluid. The tensor field $\tT_{\tU}$ connects the rigid-body motion of $6N$ particles to stress (see Appendix \ref{appendixreciprocal} for a detailed derivation). Substitution of the boundary conditions of each active particle, \eqref{bcparticle}, into \eqref{hydrodynamicforce} yields a decomposition of the hydrodynamic forces into three separate forces due to each aspect of the boundary motion: the hydrodynamic `swim' force (or thrust), 
\begin{align}
\tF_s &= \sum_i\int_{\partial\fB_i}\bu^s_i\cdot(\bn\cdot\tT_{\tU})\d S,
\label{swimforce}
\end{align}
generated by each active particle as if held fixed in an otherwise quiescent fluid; the hydrodynamic drag force on each particle 
\begin{align}
\tF_\infty &= -\sum_i\int_{\partial\fB_i}\bu^\infty\cdot(\bn\cdot\tT_{\tU})\d S,
\label{flowforce}
\end{align}
as if inactive and held fixed in a background flow; and the hydrodynamic drag due to the rigid-body motion of each particle, 
\begin{align}
\tF_d=\sum_i\int_{\partial\fB_i}(\bU_i+\bOmega_i\times\br_i)\cdot(\bn\cdot\tT_{\tU})\d S=-\tR_{\tF\tU}\cdot\tU,
\end{align}
as if inactive (passive) in an otherwise quiescent fluid. The latter is written in terms of a ($6N\times6N$) resistance tensor, which is the linear operator that gives hydrodynamic forces due to rigid-body translational/rotational velocities $\tU$ (another $6N$-dimensional vector).

Substitution of these forces into \eqref{forces} and inversion of the resistance tensor gives
\begin{align}
\tU=\tR_{\tF\tU}^{-1}\cdot\lb[\tF_{ext}+\tF_s+\tF_\infty\rb].
\label{mobility}
\end{align}
This relationship simply states that the rigid-body motion of active particles is linearly related to the forces exerted by or on those particles. The deterministic formula in \eqref{mobility} is exact and completely general; it governs the dynamics of a suspension of active (and passive) particles of arbitrary shape and activity in a general background flow. A stochastic Brownian force may also be included in the above force balance, with the associated thermal drift term that arises upon elimination of inertial degrees of freedom, 
\begin{align}
\tU=\tR_{\tF\tU}^{-1}\cdot\lb[\tF_{ext}+\tF_s+\tF_\infty+\tF_B \rb]+k_B T\bnabla\cdot\tR_{\tF\tU}^{-1}.
\label{mobility-s}
\end{align}
The vector $\tU$ now represents discrete changes in position and orientation over an interval $\Delta t$ and the Brownian force, $\tF_B = \sqrt{2k_BT/\Delta t}\tR_{\tF\tU}^{1/2}\cdot\boldsymbol{\Psi}$, where $k_B$ is the Boltzmann constant, $T$ is the fluid temperature and $\boldsymbol{\Psi}$ is a vector of standard Gaussian random variables.

Although \eqref{mobility} and \eqref{mobility-s} are exact, to compute the dynamics the tensor field $\tT_{\tU}$ would need to be found at each instant and this is prohibitively expensive for suspensions of large numbers of particles (and more so if they are changing shape). Instead, an approximate approach used in Stokesian Dynamics is to evaluate a truncated set of moments of the traction operator $\bn\cdot\tT_{\tU}$ on the surfaces of the particles $\partial\fB_i$. We outline this approach for spherical active bodies below, where the approach is particularly elegant and simplified, but the methodology can certainly be extended to anisotropic bodies \citep{claeys93a,claeys93b,claeys93c,nasouri18}.

\subsection{Spherical moments}
In order to facilitate computation we use the fact that one may write an arbitrary function on a sphere in terms of an expansion in irreducible tensors of the unit normal $\bn$ (dimensionless tensorial spherical harmonics) \citep{hess15}. In this way the deformation velocity of each active particle may be written as
\begin{align}
\bu^s(\bx \in \partial \fB_i)&=\bC^{(1)}_i+\bn\cdot\bC^{(2)}_i+\overbracket{\bn\bn}:\bC^{(3)}_i+\ldots,
\label{eqsurface}
\end{align}
where in this shorthand notation the superscript indicates the tensor order of the coefficients and the overbracket means the irreducible (or fully symmetric and traceless) part of the tensor (see the Appendix \ref{appendixtensors} for further details).

The coefficient tensors $\bC^{(n)}$ may be easily obtained by appealing to the orthogonality of the tensorial spherical harmonics
\begin{align}
\bC_i^{(n+1)} = \frac{1}{4\pi a^2}\int_{\partial\fB_i} w_n \overbracket{\bn^n}\bu^s\d S,
\end{align}
where the weight $w_n = (2n+1)!!/n!$. Hence we see the coefficient tensors are (weighted) irreducible moments of $\bu^s$. We can recast these coefficients in more familiar terms by separating symmetric and antisymmetric parts of $\bC_i^{(2)}$,
\begin{align}
\bU^s_i &=\frac{1}{4\pi a_i^2}\int_{\partial\fB_i}\bu^s\d S,\label{moments1}\\
\bOmega^s_i &= \frac{3}{8\pi a_i^4}\int_{\partial\fB_i}\br_i\times\bu^s\d S,\label{moments2}\\
\bE^s_i &=  \frac{3}{4\pi a_i^4}\int_{\partial\fB_i} \lb[\frac{1}{2}(\br_i\bu^s+\bu^s\br_i)\rb]\d S,\label{moments3}
\end{align}
to rewrite the surface velocity in familiar form \citep{swan11}
\begin{align}
\bu^s(\bx \in \partial \fB_i)&=\bU^s_i+\bOmega_i^s\times\br_i+\br_i\cdot\bE_i^s+\ldots
\end{align}

We can likewise express the background flow in terms of a moment expansion and in this way write in a consistent fashion for the disturbance field
\begin{align}
\bu'(\bx \in \partial \fB_i)&=\bU_i+\bU_i^s-\bU_i^\infty+(\bOmega_i+\bOmega_i^s-\bOmega_i^\infty)\times\br_i+\br_i\cdot(\bE_i^s-\bE_i^\infty)+\ldots
\label{velexpansion}
\end{align}
where $\bU_i^\infty$, $\bOmega_i^\infty$ and $\bE_i^\infty$ are defined as in (\ref{moments1}--\ref{moments3}) in terms of moments of the background flow $\bu^\infty$. If the background flow is linear, then $\bOmega_i^\infty=\bOmega^\infty$ and $\bE_i^\infty=\bE^\infty$ are constants everywhere in the flow (but still may be arbitrary functions of time).

Using the expansion \eqref{velexpansion} in \eqref{hydrodynamicforce} one may write the hydrodynamic forces in terms of a set of moments of the traction operator  $\bn\cdot\tT_{\tU}$ on the surfaces of the particles $\partial\fB_i$ (forming resistance tensors),
\begin{align}
\tF &=-\tR_{\tF\tU}\cdot(\tU+\tU^s-\tU^\infty)-\tR_{\tF\tE}:(\tE^s-\tE^\infty)+\ldots
\end{align}

Now using the above expression for the hydrodynamic forces, together with Newton's second law \eqref{forces}, we obtain the translational and rotational velocities of the spherical active particles
\begin{align}
\tU&=-\tU^s+\tU^\infty+\tR_{\tF\tU}^{-1}\cdot\lb[\tF_{ext}-\tR_{\tF\tE}:(\tE^s-\tE^\infty)+\ldots\rb].
\label{mobility2}
\end{align}
These equations have the same functional form as the governing equations of motion for passive particles except with active particles one takes the \textit{difference} in moments of the background flow to surface velocity, e.g. $\tE^\infty \rightarrow \tE^\infty-\tE^s$. If the particles are passive, $\bu^s=\bzero$, then we recover equations of motion for passive particles in a background flow \citep{brady88}; however, computing the far-field hydrodynamic interactions of active spherical particles is no more difficult than for passive spherical particles, assuming the velocities on the boundaries of the active particle, $\bu^s_i$, are prescribed.

If hydrodynamic interactions are completely neglected, the particles all move with their respective single particle velocities, $\tU =-\tU^s+\tU^\infty$ (higher-order moments do not contribute to self propulsion for isolated spherical particles by symmetry), and we recover the classic result for single active spheres \citep{anderson91, stone96, elfring15}. It is technically possible to devise a perfect stealth swimmer that does not disturb the surrounding fluid by setting $\bu^s = \bU^s$, for example by the jetting mechanism proposed by \citet{spagnolie10b}, but this is a pathological case, and in general higher-order moments lead to hydrodynamic interactions. We emphasize that hydrodynamic interactions due to moments of the particle activity enter in \textit{exactly} equivalent form to interactions due to moments of the background flow. For example, the leading-order change in the dynamics of the particles due to hydrodynamic interactions is given by $\tU+\tU^s-\tU^\infty=-\tR_{\tF\tU}^{-1}\cdot\tR_{\tF\tE}:(\tE^s-\tE^\infty)$ where the resistance tensors $\tR_{\tF\tU}^{-1}\cdot\tR_{\tF\tE}$ act to couple the particles in precisely the same fashion for active particles as passive particles. In appendix \ref{app:dilute} we given the leading order hydrodynamic interactions (a dilute approximation) in the mobility formulation more commonly employed in the literature.

We see that the leading order hydrodynamic interactions due to activity are given by the symmetric first moment of activity, $\bE^s_i$, of each active particle. This is not a surprise as the term $\bE^s$ sets the active component of the stresslet $\bS$ \citep{ishikawa06,nasouri18} of individual spherical active particles where
\begin{align}
\bS = \frac{1}{2}\int_{\partial\fB}[\br\bsigma\cdot\bn+\bsigma\cdot\bn\br -2\eta(\bu\bn+\bn\bu) ] \d S = \frac{20\pi\eta a^3}{3}\lb(\bE^\infty-\bE^s\rb).
\end{align}
The `active strain-rate', $\bE^s$, can be zero, but then the leading-order term will generally arise at the second-moment level for self-motile active particles (that is, ones with non-zero surface averaged velocity). This is the case for so-called neutral squirmers (see section \ref{sec:squirmers} below) or symmetric phoretic particles \citep{michelin14}.

These equations, above all, simply reflect the linear relationship between velocity and force moments. Using still more compact notation for all disturbance velocity moments $\fU' = [\tU+\tU^s-\tU^\infty, \tE^s-\tE^\infty,\ldots]^\top$ and hydrodynamic force moments $\fF = [\tF, \tS,\ldots]^\top$ we may write, more generally, the linear relationship
\begin{align}
\fF=-\fR\cdot\fU',
\label{grandmoments}
\end{align}
where $\fR$ is the grand resistance tensor, an (unbounded) linear operator that maps velocity moments to force moments. In this notation the hydrodynamic force is compactly written as
\begin{align}
\tF = -\fR_{\tF\fU}\cdot\fU'.
\end{align}

In order to capture the dynamics of active particles, we seek an effective and efficient way to form $\fR_{\tF\fU}$. The grand resistance tensor is a purely geometric operator, depending only on the position (and orientation if they were anisotropic) of each active particle \citep{happel65}. Perhaps less obvious, is that the grand resistance tensor does not depend on the prescribed surface activity of the particles, and is thus identical to the case when they are passive. This also applies to particles in a bounded geometry---$\fR$ is a function of geometry only \citep{swan07,swan10}. We have assumed here that the surface activity of the particles is prescribed; however, in reality the surface activity may depend on the traction on the boundary, as it would, for example, for biological particles that have power-limited surface actuation, but the linear relationship between force moments and velocity moments makes it straightforward to alternatively prescribe force moments \citep{swan11}.

We focus here on spherical particles as is common in the literature for colloidal suspensions; however, the method described above can be readily generalized to other geometries. We formed moments of forces and velocities by projection onto tensorial spherical harmonics, but for other geometries a more suitable basis for the vector fields on the particle surfaces, $\partial\fB_i$, would be used. Alternatively, and more generally, one may perform Taylor series expansion of the boundary integral equations about the center of each particle, which naturally projects tractions onto force moments for particles of arbitrary geometry (see recent work by \citet{swan11} and \citet{nasouri18} for details of this method applied to active particles). A problem with this approach is that the particle activity $\bu^s$ might only be defined on the particle surfaces (in the form of surface slip as in section \ref{sec:squirmers}) but this difficulty can be ameliorated by lifting $\bu^s$ to a suitably continuous function defined in $\mathbb{R}^3$. Despite this complication, fundamentally, the linear relationship between velocity and force moments remains, regardless of geometry.

\subsection{Squirmers}\label{sec:squirmers}
A squirmer is a spherical particle whose surface slip velocity is tangential to the surface \citep{pedley16}. Most often the slip velocity is taken to be axisymmetric such that $\bu^s = f(\phi)\be_\phi+g(\phi)\be_\theta$, where $\phi$ is the (polar) angle between the axis of symmetry of the particle $\bp$ and the surface normal $\bn$, while the azimuthal direction $\be_\theta = \bn\times\be_\phi$. A purely tangential slip velocity is of course an idealization, but one that arises quite naturally, for example in the limit of small amplitude deformations that are projected onto a time-averaged spherical manifold \citep{lighthill52,blake71}, or as the outer solution of phoretic flow due to chemical concentrations confined to a thin layer near the sphere surface \citep{anderson89, golestanian05}. The slip velocity is typically written as an expansion in Legendre polynomials
\begin{align}
\bu^s &= -\sum_n B_n \frac{2}{n(n+1)}P_n^1 (\bp\cdot\bn)\be_\phi -\sum_n C_n \frac{2}{n(n+1)}P_n^1 (\bp\cdot\bn)\be_\theta
\end{align}
where $P_n^1$ is the first associated Legendre polynomial of degree $n$ \citep{pak14} ($C_n\rightarrow C_n \frac{n(n+1)}{2a^{n+1}}$ for their coefficients). The polar slip coefficients, $B_n$, are often called `squirming' modes while the azimuthal slip is not often considered but can lead to particle spin for instance \citep{pak14}. Recasting the slip velocity in terms of irreducible tensors of the surface normal such that
\begin{align}
\bu^s(\bx \in \partial \fB)&=\bU^s+\bOmega^s\times\br+\br\cdot\bE^s+\overbracket{\br^2}:\bB^s+\overbracket{\br^3}\odot\bC^s+\ldots \ ,
\label{velocitymoments}
\end{align}
we obtain
\begin{align}
\bU^s &= -\frac{2}{3}B_1 \bp, \\
\bOmega^s &= \frac{1}{a^3}C_1\bp,\\
a\bE^s &= -\frac{3}{5}B_2 \overbracket{\bp\bp},\\
a^2\bB^s &= B_1\bDelta^{2}\cdot\bp-\frac{5}{7}B_3\overbracket{\bp\bp\bp}-C_2(\bDelta^2\cdot\bp)\times\bp,\\
a^3\bC^s &= B_2\bDelta^{3}:\overbracket{\bp\bp}-\frac{35}{36}B_4\overbracket{\bp\bp\bp\bp}-\frac{5}{4}C_3(\bDelta^3:\overbracket{\bp\bp})\times\bp,
\end{align}
where $\bDelta^{n}$ is an isotropic $2n$ order tensor that when applied on a tensor of rank $n$, projects onto the symmetric traceless part of that tensor (see Appendix \ref{appendixtensors} for further details). By symmetry each coefficient is necessarily composed only of products of the particle director $\bp$. We see that the swimming speed is given by the first squirming mode $B_1$, $\bU=-\bU^s=(2/3)B_1\bp$ for an isolated squirmer, but note that the first mode also contributes a higher-order term to hydrodynamic interactions between particles embedded in $\bB^s$. The stresslet due to surface activity of a particle is given by the second squirming mode, $\bS = 4\pi \eta a^2 B_2\overbracket{\bp\bp}$. This determines if a squirmer is a pusher or a puller, but can easily be zero---a so-called neutral squirmer---and in that case the leading-order term contributing to hydrodynamic interactions is necessarily given by $B_1$ (and $B_3$ if nonzero). Azimuthal slip naturally leads to rotation given by the $C_1$ mode, $\bOmega=-\bOmega^s = -(C_1/a^3)\bp$ for an isolated squirmer, while the $C_2$ mode leads to a rotlet dipole contribution in the far-field \citep{pak14}.

\begin{figure}
\center
\includegraphics[width = 0.35\textwidth]{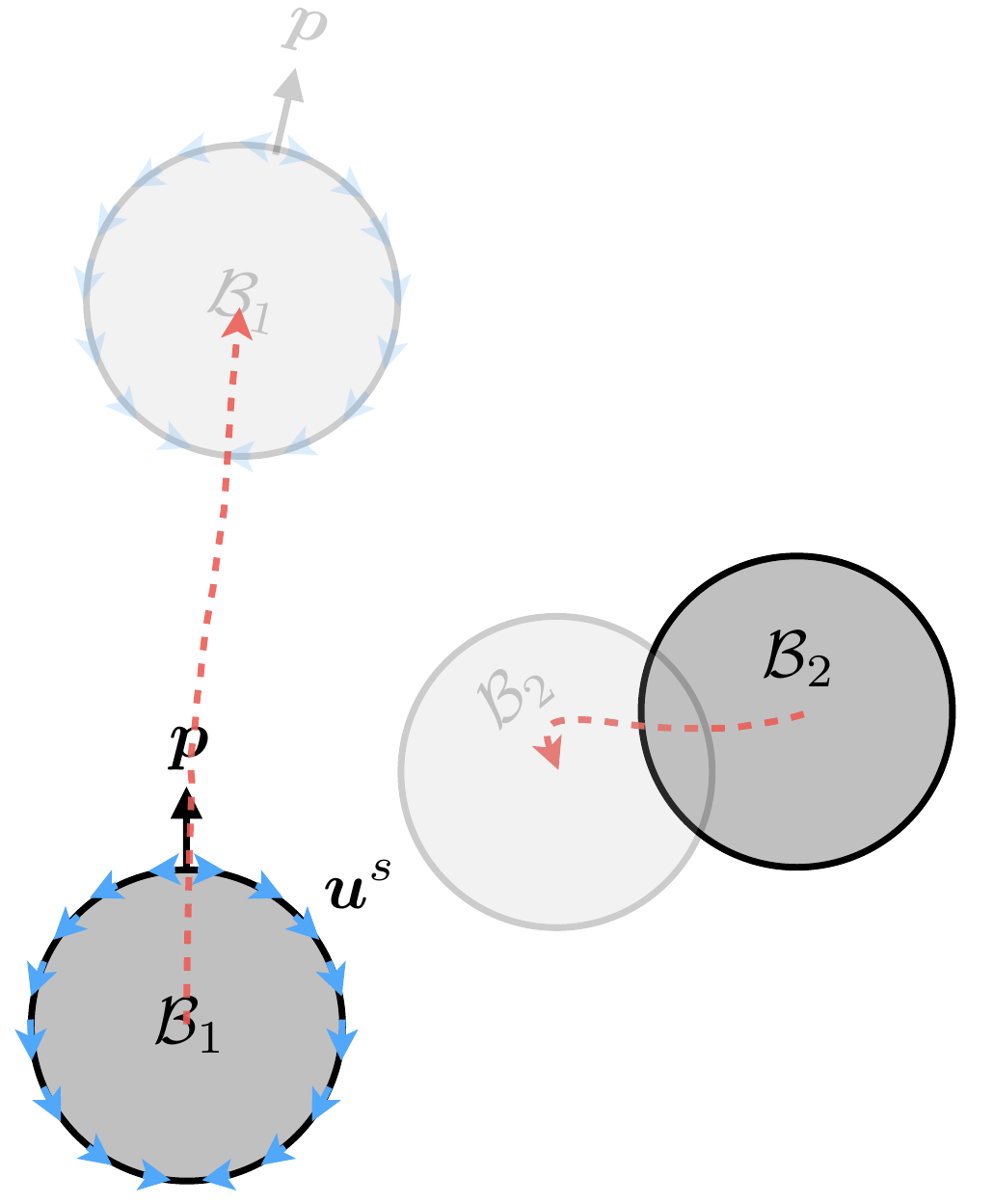}
\caption{Trajectory of a (pusher) active particle (labeled $\fB_1$), in the presence of a passive particle (labeled $\fB_2$).}
\label{fig:2particle_traj}
\end{figure}

As an example of the framework developed here, consider an active squirmer particle, labeled $\fB_1$, in the presence of a freely suspended passive sphere, labeled $\fB_2$, as shown in figure \ref{fig:2particle_traj}.
Using \eqref{mobility2} we obtain the velocities of the two particles in terms of moments of the surface activity of the active particle
\begin{align}
\tU_1 &= -\tU_1^s - (\tM_{\tU\tF}^{11}\cdot\tR_{\tF\tE}^{11}+ \tM_{\tU\tF}^{12}\cdot\tR_{\tF\tE}^{21}):\bE_1^s+\ldots \label{2particle1}\\
\tU_2 &= - (\tM_{\tU\tF}^{21}\cdot\tR_{\tF\tE}^{11} + \tM_{\tU\tF}^{22}\cdot\tR_{\tF\tE}^{21}):\bE_1^s+\ldots \label{2particle2}
\end{align}
where the superscripts, for example $\tR_{\tF\tE}^{\alpha\beta}$, indicate the linear relationship between particle $\alpha$ and particle $\beta$, while $\tM_{\tU\tF} = \tR_{\tF\tU}^{-1}$. The first term on the right-hand side of \eqref{2particle1} represents the self-propulsion of the active particle while the second term represents the change in the velocity due to hydrodynamic interactions induced by the surface strain-rate of the active particle $\bE^s_1$ (and higher-order moments). Hydrodynamic interactions also induce the motion of the passive particle. In essence, the moments of $\bu^s$ on the active particle result in a `swim' force on \textit{both} particles which must be then balanced by drag due to rigid-body motion. The trajectories of both active and passive particles are illustrated in figure \ref{fig:2particle_traj}.

\subsection{Assemblies}
As detailed by \citet{swan11}, assemblies of active (or passive) particles can easily be dealt with within the Stokesian Dynamics framework, and in what follows we outline the presentation given in that work.

A set of particles may be constrained to move as a rigid body, namely particle $\alpha$ in a rigid assembly $A$, will move as
\begin{align}
\bU_{\alpha} &= \bU_A+\bOmega_A\times(\bx_\alpha-\bx_A),\\
\bOmega_{\alpha} &= \bOmega_A.
\end{align}
where $\bx_A$ is a convenient point on the assembly. Following the notation in \citet{swan11}, this may be compactly written in terms of $6-$dimensional vectors, $\tU_\alpha = \bSigma_{\alpha A}^{\top}\cdot\tU_A$, where $\bSigma_{\alpha A}^{\top}$ projects the translational and rotational velocity of the assembly onto particle $\alpha$. The rigid-body translational and rotational velocities of all $N$ particles in an assembly may then be written in terms of $6N-$dimensional vectors and tensors as
\begin{align}
\tU = \bSigma_{A}^{\top}\cdot\tU_A.
\label{assembly}
\end{align}

The forces and torques that enforce the rigid constraints on the assembly, $\tF_c$, must be included in the sum of forces on the particles,
\begin{align}
\tF+\tF_c+\tF_{ext}=\tzero.
\end{align}
These constraint forces are internal forces, and as such exert no net force or torque on the assembly. This may be written as
\begin{align}
\bSigma_{A}\cdot\tF_c = \tzero,
\end{align}
where the operator $\bSigma_{A}$, the transpose of the projection above, sums forces and torques (about $\bx_A$) on the assembly. In this way the force balance on the assembly is
\begin{align}
\bSigma_{A}\cdot\tF+\bSigma_{A}\cdot\tF_{ext}=\tzero.
\end{align}
Substitution of the relevant hydrodynamic forces and the kinematic constraint in \eqref{assembly} into this force balance leads to the rigid-body motion of the assembly given by
\begin{align}
\tU_A=\lb[\bSigma_{A}\cdot\tR_{\tF\tU}\cdot\bSigma_A^\top\rb]^{-1}\cdot\bSigma_{A}\cdot\lb(\tF_{ext}+\tF_s+\tF_\infty\rb),
\label{assembly2}
\end{align}
where $\bSigma_{A}\cdot\tR_{\tF\tU}\cdot\bSigma_A^\top=\tR_{\tF\tU}^A$ is the hydrodynamic resistance of the assembly. Equation \eqref{assembly2} is an exact description of the dynamics of an assembly of active or passive particles, no approximation has yet been made. In particular, we note that while \eqref{assembly2} yields the instantaneous rigid-body motion of the assembly, it does not mean the assembly cannot deform. Indeed, through the prescription of the activity of each particle, by way of $\bu^s$, we may construct an assembly of virtually any shape and kinematics. This approach is also straightforwardly extended to multiple assemblies through an extended operator $\bSigma$ that sums forces on each assembly as shown by \citet{swan11}. As discussed above, a natural method of solution is to use Stokesian Dynamics to resolve hydrodynamic forces as a truncated set of moments.

\begin{figure}
\center
\includegraphics[width = 0.5\textwidth]{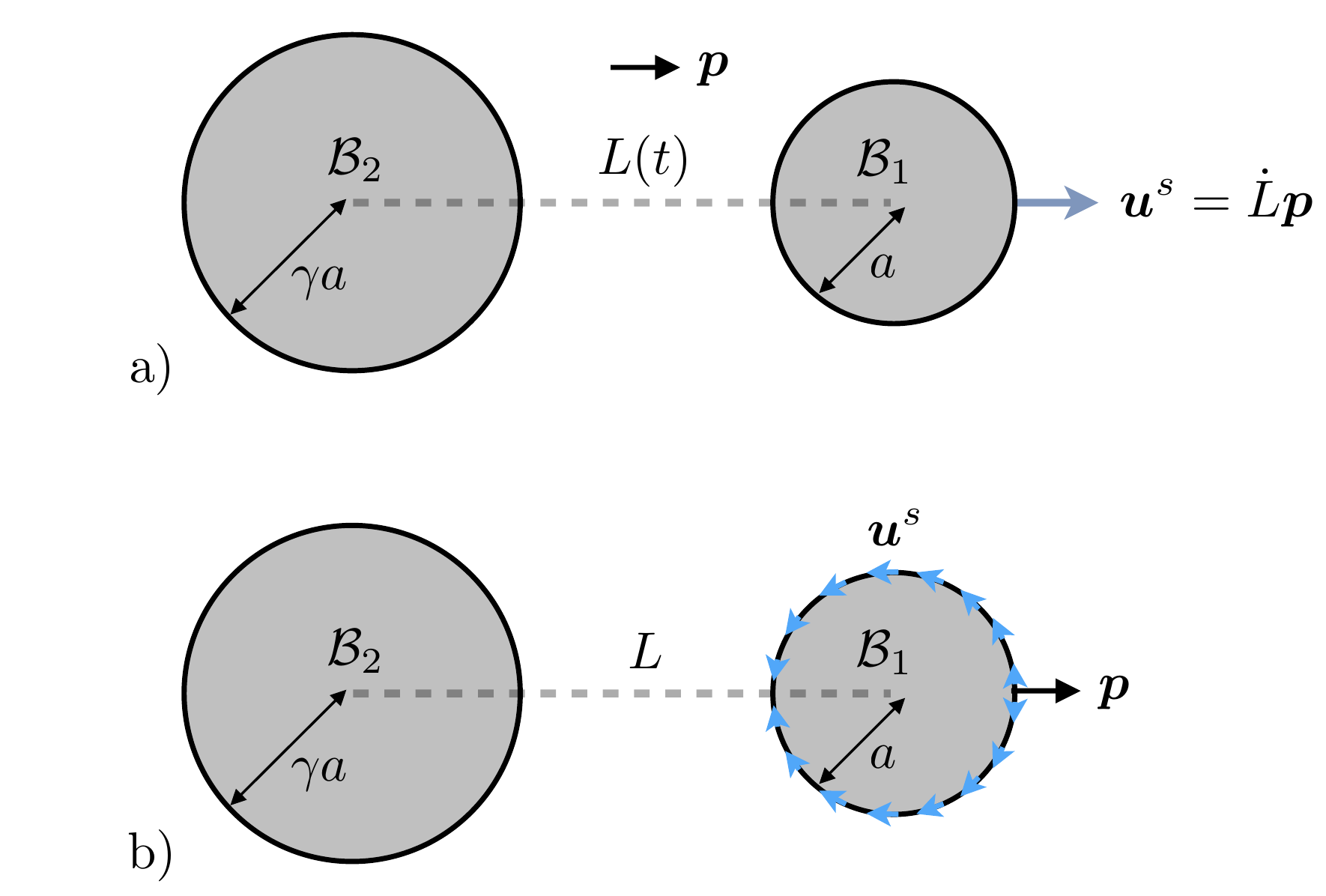}
\caption{a) Schematic of a reciprocal dumbbell swimmer. b) Schematic of a dumbbell squirmer swimmer.}
\label{fig:2particle}
\end{figure}

As an illustrative example of a deforming assembly, consider a simple reciprocal two-sphere  (or dumbbell) swimmer (see figure \ref{fig:2particle}a). In this model swimmer, two spheres labeled $\fB_1$ and $\fB_2$ ($\fB_A = \fB_1\cup\fB_2$), of radius $a$ and $\lambda a$ respectively, have a prescribed distance between their centers, $L(t)$, that periodically changes in time.
We describe the shape change of this swimmer as motion of sphere $\fB_1$ relative to sphere $\fB_2$, in this way $\bu^s$ is nonzero only on $\fB_1$. Written in terms of an expansion in moments as in \eqref{velocitymoments}, $\bu^s(\bx\in\fB_1)=\bU_1^s=\dot{L}\bp$ with all other terms exactly zero, while $\bu^s(\bx\in\fB_2)=\bzero$. The total velocity of $\fB_2$ is then due solely to the rigid-body motion of the assembly $\bu(\bx\in\fB_2)=\bU_A$, while $\fB_1$ has an additional component due to shape change $\bu(\bx\in\fB_1)=\bU_A+\bU_1^s$. By symmetry this swimmer does not rotate, $\bOmega_A=\bzero$. The choice of reference is not unique and affects what is delineated as rigid-body motion versus shape change at any particular instant; however, typically we are concerned with the time-averaged motion of the body which is invariant to the choice of reference for periodic gaits, and the flexibility allows one take advantage of simplifications implied by a particular choice.

Due to the lack or rotation or torque, only the force-velocity resistance tensor of the assembly, $\bR_{FU}^A$, and the linear operator that gives stress due to translation, $\tT_U$, are required. Substitution of the swim force \eqref{swimforce}, into \eqref{assembly2} and simplification leads to
\begin{align}
\bU_A &= -(\bR_{FU}^{A})^{-1}\cdot\lb(\bR_{FU}^{11}+\bR_{FU}^{21}\rb)\cdot\bU_1^s,
\end{align}
where the resistance tensors, $\bR_{FU}^{11}+\bR_{FU}^{21}$ and $\bR_{FU}^{A}=\bR_{FU}^{11}+\bR_{FU}^{12}+\bR_{FU}^{21}+\bR_{FU}^{22}$, are functions of the length, $L(t)$, and hence depend on time. We may further simplify by noting that the propulsive force and velocity will be collinear with the axis of symmetry and so only a scalar coefficient for each resistance is required. A symmetric swimmer with $\lambda=1$ has $\bU_A =-\frac{1}{2}\bU_1^s= -\frac{1}{2}\dot{L}\,\bp$, so the dumbbell   moves oppositely of the deformation with half the speed, as expected. This reciprocal motion clearly leads to zero net displacement over a period when $L(t)$ is periodic. Less obvious, but also true, is that this holds for any $\lambda$, by the scallop theorem \citep{purcell77}. 

The previous example was particularly straightforward because $\bu^s$ was uniform hence only the zeroth moment, $\bU^s$, was nonzero. In contrast, consider a dumbbell swimmer with a fixed length, $L=const.$, but where sphere $\fB_1$ is a squirmer particle (see figure \ref{fig:2particle}b), namely $\bu^s(\bx\in\fB_1) = \bU^s_1+\br\cdot\bE_1^s+\ldots$, with the moments of the surface velocity given by the squirming modes. In this case, the velocity of the assembly is given by
\begin{align}
\bU_A &= -(\bR_{FU}^{A})^{-1}\cdot\lb[\lb(\bR_{FU}^{11}+\bR_{FU}^{21}\rb)\cdot\bU_1^s+\lb(\bR_{FE}^{11}+\bR_{FE}^{21}\rb):\bE_1^s\ldots\rb],
\end{align}
and when the spheres are equal in size, $\lambda = 1$, we have simply $\bU_A = -\frac{1}{2}\bU_1^s-(\bR_{FU}^{A})^{-1}\cdot\lb[\bR_{FE}^{1}:\bE_1^s+\ldots\rb]$. Note that this swimmer can self propel even when $\bU_1^s=\bzero$ due to hydrodynamic interactions with the second sphere.

\section{Stokesian Dynamics}\label{sec:stokesiandynamics}
The configuration dependent $N$-body resistance tensors may be formed indirectly by first constructing the grand mobility tensor, $\fM=\fR^{-1}$. In this approach, one takes irreducible moments of the velocity field, as given by the boundary integral equation, over the surfaces of all the particles yielding Fax\'en's laws for the velocity moments of the active particles \citep{batchelor72}. If the boundary integral equations are also expanded in irreducible moments \citep{durlofsky87}, we obtain a linear relationship between force and velocity moments
\begin{align}
\fU' = -\fM\cdot\fF.
\end{align}
For active particles, the force moments contain contributions from the double-layer kernel due to the surface activity (see appendix for details). The grand mobility tensor is then inverted to obtain the grand resistance tensor $\fR=\fM^{-1}$ thereby summing many-body hydrodynamic interactions among the particles \citep{durlofsky87}. In principle, to capture near-field lubrication effects, the entire unbounded set of moments would need to be computed, and in practice this is unfeasible. The coupling between the $m^{\text{th}}$-moment of velocity and $n^{\text{th}}$-moment of force scales as $r^{-(1+m+n)}$ , and so higher-order moments decay quite quickly with separation distance $r$ between two particles and a reasonable and common far-field approximation of the mobility is to truncate at the first moment level and we label this truncated mobility $\fM^{ff}$ \citep{swan11}. This level of approximation is inappropriate for particles that are nearly touching and the compromise used within Stokesian Dynamics is to use a mixed asymptotic approach wherein close interactions are separately computed using pairwise exact solutions \citep{durlofsky87}. In this manner, the hydrodynamic forces on the particles are decomposed,
\begin{align}
\tF = \tF^{ff}+\tF^{2B,exact}-\tF^{2B,ff},
\end{align}
into far-field interactions between many bodies, $\tF^{ff}$ and two-body interactions computed exactly for nearby bodies, $\tF^{2B,exact}$. Note that the last term arises because the far-field interactions must be removed between any two bodies where the interactions are computed exactly to avoid double counting.

To render exact solutions for active two-body hydrodynamic interactions one needs to obtain the tensor field $\tT_{\tU}$ from the two particle rigid-body motion problem. The general passive two-sphere problem for arbitrary separations may be constructed from a basis of four simplified two-sphere problems that have all be solved in the literature and are nicely summarized by \citet{sharifi-mood16} and \citet{papavassiliou17} in the context of two-body interactions between diffusiophoretic Janus particles and spherical squirmers respectively. Asymptotic solutions for lubrication interactions, which are valid strictly only when the particles are very close, may alternatively be used and are given by \citet{ishikawa06} for spherical squirmers.

It is important to note that, unlike passive particles in a linear background flow, active particles can have higher-order velocity moments due to surface activity. In section \ref{sec:squirmers}, we showed that even two-mode squirmer particles contribute third and fourth order velocity moments. For far-field interactions, these higher-order moments may not be significant due to the decay of the associated flow disturbances. However, for near-field interactions there is no rationale, other than the convergence of the series of tensorial spherical harmonics, to discard the contributions of higher-order velocity moments in the swim force for near-field interactions. If higher-order moments are nonetheless discarded then the approach for active particles is virtually identical to that of passive particles in Stokesian Dynamics; the dynamics are given by \eqref{mobility2} and the resistance tensor used is modified to include both exact two-body interactions  $\fR^{2B,exact}$, and a truncation of the moment expansion valid for far-field interactions, $(\fM^{ff})^{-1}$, such that
\begin{align}
\fR = (\fM^{ff})^{-1}+\fR^{2B,exact}-\fR^{2B,ff},
\label{totalresistance}
\end{align}
where the two-body interactions that are captured by the near-field approach must be subtracted off in the far-field solution to avoid double counting \citep{swan11}. 

More accurately, the exact two-body swim force contributions may be computed entirely separately, as done by \cite{ishikawa06}, by integrating \eqref{swimforce} directly. In this way \eqref{mobility} is written
\begin{align}
\tU&=\tU^\infty+\tR_{\tF\tU}^{-1}\cdot\lb[\tF_{ext}+\tR_{\tF\tE}:\tE^\infty+\ldots\rb]\nonumber\\
&\quad+\tR_{\tF\tU}^{-1}\cdot\lb[\tF_{s}^{2B,exact}-\tR'_{\tF\tU}:\tU^s-\tR'_{\tF\tE}:\tE^s+\ldots\rb].
\label{mobility3}
\end{align}
Here the terms on the first line represent the dynamics of passive spheres exactly as in conventional Stokesian Dynamics, while the second line accounts for the contribution of activity, separated into two-body and far-field contributions. The primed resistance tensors only contribute far-field interactions
\begin{align}
\fR' = (\fM^{ff})^{-1}-\fR^{2B,ff}.
\end{align}
If Brownian motion is included then it is $\tR_{\tF\tU}$ from the total resistance \eqref{totalresistance} that sets the magnitude of the Brownian force.

\subsection{Infinite suspensions}
The method described above was for a finite system of $N$ active particles for which the fluid can be assumed to decay in the far field. For an infinite or periodic suspension of particles (active or passive) no such assumption can be made and indeed naively extending $N\rightarrow\infty$ leads to divergent integrals, a problem that plagued the earlier suspension literature \citep{batchelor72}. \citet{brady88a} adapted the method of \citet{obrien79} wherein the fluid domain for a set of particles is bounded by a large macroscopic surface over which suspension averages can be performed. Specifically, $\bU^\infty$, $\bOmega^\infty$, and $\bE^\infty$ become the average values of the suspension --- particle plus fluid. Individual particle motion is then relative to the volume averaged quantities. Suspension-averaged terms serve to regularize the formulas leading to absolutely convergent expressions for fluid and particle velocities. Periodic boundary conditions may then be easily employed and as \citet{brady88a} showed, the far-field mobility matrix, $\fM^{ff}$, may be simply replaced by the appropriated Ewald-summed mobility matrix, $\fM^{ff*}$. As discussed above, the mobility matrix is unchanged if particles are active or passive, only the force and velocity moments are altered by activity, and mobility is unchanged whether or not the suspension averaged quantities are nonzero. Therefore, the Ewald-summed mobility matrix used for periodic passive suspensions is unchanged for active suspensions \citep{ishikawa08}. It is important to note that for self propulsion there is no net volume displacement of material as the body moves: as the body advances, an equal volume of fluid moves in the opposite direction. In contrast, a body moving in response to an external force drags fluid along with it, and to have no net flux of mass, an external pressure gradient must be imposed.

\subsection{Accelerated methods}
Since the original development of the Stokesian Dynamics method \citep{durlofsky87,brady88a}, which naively requires $O(N^3)$ computations due to the inversion of the far-field mobility matrix, there have been several numerical implementations that have improved the algorithmic efficiency of the method. These include an $O(N \ln N)$ deterministic Accelerated Stokesian Dynamics method \citep{sierou01}, an $O(N^{1.25} \ln N)$ Brownian Accelerated Stokesian Dynamics method \citep{banchio03}, an $O(N \ln N)$ Spectral Ewald Accelerated Stokesian Dynamics method \citep{wang16}, and recently an $O(N)$ Fast Stokesian Dynamics method \citep{fiore19}. In principle, because the structure of the Stokesian Dynamics framework remains unchanged between passive and active particles, any of these approaches may be used to simulate active suspensions with minor modification. Indeed, the recent Fast Stokesian Dynamics method utilizes an imposed Brownian `slip' velocity in order to obtain the stochastic rigid-body motion of passive Brownian particles \citep{fiore19}.

\section{Conclusions}
In this work, we have given a detailed exact theoretical description of the dynamics suspensions of active particles in fluids in the absence of inertia, including full hydrodynamic interactions among particles. We argue that, as is done for passive particles, hydrodynamic interactions are ideally separated into near-field forces and far-field forces, with the latter expanded in a truncated set of moments. The resulting mathematical structure of the dynamical equations remains virtually unchanged between passive and active particles save for the addition of velocity moments due to particle activity. Because of this, any implementation of the Stokesian Dynamics method for passive particles, may be simply and easily modified for use with active particles. Moreover, we believe this mathematical structure provides an ideal formalism for theoretical analysis of hydrodynamic interactions in active matter much as it has for passive suspensions.
\acknowledgements
Gwynn Elfring acknowledges the hospitality of the Division of Chemistry and Chemical Engineering at the California Institute of Technology during a sabbatical stay, supported by a UBC Killam Research Fellowship, that served as a formative period of this work.


\appendix
\section{Tensorial spherical harmonics}\label{appendixtensors}
The (n-adic) tensorial spherical harmonics are a set of tensors composed of irreducible products of the unit normal on a sphere \citep{brenner64c, hess15}
\begin{align}
\overbracket{\bn^2}&=\overbracket{\bn\bn}=\bn\bn-\frac{1}{3}\bI,\\
\overbracket{\bn^3}&=\overbracket{\bn\bn\bn} = \bn\bn\bn - \frac{1}{5}\lb(\bI\bn+\bn\bI+\prescript{\top}{}(\bn\bI)]\rb),\\
\lb[\overbracket{\bn^4}\rb]_{ijkl} &= n_in_jn_kn_l - \frac{1}{7}\lb(n_in_j\delta_{kl}+n_i\delta_{jk}n_l+\delta_{ij}n_kn_l+\delta_{il}n_jn_k+\delta_{ik}n_jn_l+n_i\delta_{jl}n_k\rb),\nonumber\\
&\quad +\frac{1}{35}\lb(\delta_{ij}\delta_{kl}+\delta_{ik}\delta_{jl}+\delta_{il}\delta_{jk}\rb)\\
\overbracket{\bn^n}&=\frac{(-1)^n}{(2n-1)!!}a^{n+1}\bnabla^{n}\lb(\frac{1}{r}\rb)_{r=a}.
\end{align}
where we use notation similar to \citet{brenner64d} such that $\bnabla^2 \equiv \bnabla\bnabla$ (distinct from $\nabla^2=\bnabla\cdot\bnabla$), the overbracket indicates the irreducible (or fully symmetric and traceless) part of the tensor, and such that the transpose applies to the two adjacent indices (for example $\prescript{\top}{}\ba\bb\bc=\bb\ba\bc$ and $\ba\bb\bc^\top=\ba\bc\bb$).

The tensorial spherical harmonics are orthogonal, with the relationship
\begin{align}
\frac{1}{4\pi a^2}\int w_n \overbracket{\bn^m}  \overbracket{\bn^n} \d S = \delta_{mn}\bDelta^n,
\label{ortho}
\end{align}
where the weight 
\begin{align}
w_n = \frac{(2n+1)!!}{n!}.
\end{align}
The isotropic tensor $\bDelta^n$ is a $2n$-order tensor that projects an $n$-order tensor into its symmetric irreducible form \citep{hess15}, i.e. for the $n$-order tensor $\bA$, $\bDelta^{n}\odot\bA=\overbracket{\bA}$ where $\odot$ is a complete tensor contraction. The first several symmetrizing tensors are
\begin{align}
\Delta^0&=1,\\
\Delta^1_{ii'}&=\delta_{ii'},\\
\Delta^2_{iji'j'}&=\frac{1}{2}\lb(\delta_{ii'}\delta_{jj'}+\delta_{ij'}\delta_{ji'}\rb)-\frac{1}{3}\delta_{ij}\delta_{i'j'}\\
\Delta^3_{ijki'j'k'}&=\frac{1}{6}\lb(\delta_{ii'}\delta_{jj'}\delta_{kk'}+\delta_{ii'}\delta_{jk'}\delta_{kj'}+\delta_{ij'}\delta_{ji'}\delta_{kk'}+\delta_{ij'}\delta_{jk'}\delta_{ki'}+\delta_{ik'}\delta_{ji'}\delta_{kj'}+\delta_{ik'}\delta_{jj'}\delta_{ki'}\rb)\nonumber\\
&\quad-\frac{1}{15}\big\{\lb(\delta_{i'j'}\delta_{kk'}+\delta_{i'k'}\delta_{kj'}+\delta_{j'k'}\delta_{ki'}\rb)\delta_{ij}+\lb(\delta_{i'j'}\delta_{ik'}+\delta_{i'k'}\delta_{ij'}+\delta_{j'k'}\delta_{ii'}\rb)\delta_{jk}\nonumber\\
&\qquad+\lb(\delta_{i'j'}\delta_{jk'}+\delta_{i'k'}\delta_{jj'}+\delta_{j'k'}\delta_{ji'}\rb)\delta_{ik}\big\},
\end{align}
where the primed indices are distinct from the unprimed ones.

\section{A reciprocal theorem for active particles}\label{appendixreciprocal}
We derive here an equation for the hydrodynamic forces on active particles as shown in \eqref{hydrodynamicforce} using the reciprocal theorem (see the excellent review by \citet{masoud19}). The presentation here largely follows that found in our other work \citep{elfring14,elfring17} and but also found elsewhere \citep{papavassiliou15}. Consider $N$ active free particles $\fB_i$ with surfaces $\partial \fB_i$, where $i\in [1,N]$, with boundary conditions $\bu(\bx\in \partial\fB_i) = \bU_i+\bOmega_i\times\br_i+\bu_i^s$ immersed in a background flow $\bu^\infty$ (note that $\bu^\infty$ describes the background flow without the presence of the particle). As an auxiliary problem, here denoted by a hat, consider $N$ bodies of the same instantaneous shape undergoing rigid-body motion, $\buh(\bx\in \partial\fB_i) = \bUh_i+\bOmegah_i\times\br_i$ in a quiescent fluid (although not necessary, we take the fluids to have equal viscosity). All flow fields are incompressible and we neglect inertia in the fluid so that we may write
\begin{align}
\bnabla\cdot\lb(\bsigma'\cdot\buh-\bsigmah\cdot\bu'\rb)=\bzero,
\end{align}
where we a define disturbance flow $\bu' = \bu -\bu^\infty$ and disturbance stress $\bsigma'=\bsigma-\bsigma^\infty$

We now integrate over a (sufficiently extended) volume of fluid exterior to $\fB$ and apply the divergence theorem. Provided the fields, $\bu'$ and $\bsigma'$, decay appropriately in the far-field \citep{leal80}, we obtain
\begin{align}
\sum_i\int_{\partial\fB_i}\bn\cdot\bsigma'\cdot\buh\d S=\sum_i\int_{\partial\fB_i}\bn\cdot\bsigmah\cdot\bu'\d S,
\end{align}
Here, $\bn$ is the normal to the surface, $\partial \fB_i$, pointing into the fluid.  This is a statement of the equality of the virtual power of the motions of $\partial\fB_i$ between $\bu$ and $\buh$ \citep{happel65}.

Applying the boundary conditions for the rigid-body motion of the particles in the auxiliary problem we obtain
\begin{align}\label{total1}
\sum_i\lb[\bF_i\cdot\bUh_i+\bL_i\cdot\bOmegah_i\rb]
=\sum_i\int_{\partial\fB_i}\bn\cdot\bsigmah\cdot\bu'\d S,
\end{align}
where the hydrodynamic force and torque on the active particles are respectively
\begin{align}
\bF_i &= \int_{\partial\fB_i}\bn\cdot\bsigma'\d S,\\
\bL_i &= \int_{\partial\fB_i}\br_i\times(\bn\cdot\bsigma')\d S,
\end{align}
where we drop the primes because the background flow is force- and torque-free.

Introducing a more compact notation, where
\begin{align}
\tF &=\lb[\bF_1, \ \bL_1, \ \bF_2, \ \bL_2, \ \ldots\rb],\\
\tUh &=\lb[\bUh_1, \ \bOmegah_1, \ \bUh_2, \ \bOmegah_2, \ \ldots\rb],
\end{align}
we obtain
\begin{align}
\tF\cdot\tUh=\sum_i\int_{\partial\fB_i}\bn\cdot\bsigmah\cdot\bu'\d S.\label{totalnot1}
\end{align} 
Now, by linearity, we may write $\bsigmah = \tT_{\tU}\cdot\tUh$ and substitution into \eqref{totalnot1}, upon discarding the abritrary $\tUh$, leads to equation \eqref{hydrodynamicforce} for the hydrodynamic forces and torques on all $N$ active particles
\begin{align*}
\tF=\sum_i\int_{\partial\fB}\bu_i'\cdot(\bn\cdot\tT_{\tU})\d S.
\end{align*}

This derivation extends naturally to higher-order force moments by taking the auxiliary problem to be rigid-body motion in an arbitrary background flow, represented as a series expansion \citep{elfring17,nasouri18}.

\section{Expansion of the boundary integral equation}\label{appendixBI}
We derive here the grand mobility relationship between velocity moments and force moments by means of a Galerkin projection onto tensorial spherical harmonics \citep{singh15, fiore18}. Consider the boundary integral equation for a suspension of active particles
\begin{align}
\bu(\bx) -\bu^\infty(\bx) =  -\sum_i\int_{\partial\fB_i}\lb[\bG(\bx,\by)\cdot\bf(\by)+\bu(\by)\cdot\bT(\bx,\by)\cdot\bn(\by)\rb]\d S(\by),
\end{align}
where
\begin{align}
\bG(\bx,\by) = \frac{1}{8\pi\eta}\lb(\frac{\bI}{\lb|\bx-\by\rb|}+\frac{(\bx-\by)(\bx-\by)}{\lb|\bx-\by\rb|^3}\rb),\\
\bT(\bx,\by) = -\frac{3}{4\pi}\frac{(\bx-\by)(\bx-\by)(\bx-\by)}{\lb|\bx-\by\rb|^5}.
\end{align}
The velocities and tractions on each particle are now expressed in terms of expansions in tensorial spherical harmonics
\begin{align}
\bu(\by \in \partial \fB_i)&=\bU_i+\bU^s_i+(\bOmega_i+\bOmega_i^s)\times\br_i+\br_i\cdot\bE_i^s+\ldots,\\
\bf(\by\in\partial\fB_i) &= \frac{1}{4\pi a_i^2}\bF_i+\frac{3}{8\pi a_i^3}\bL_i\times\bn+\frac{3}{4\pi a_i^3}\bn\cdot\bSt_i+\ldots
\end{align}
where
\begin{align}
\bF_i &=\int_{\partial\fB_i}\bf\d S,\\
\bL_i &= \int_{\partial\fB_i}\br_i\times\bf\d S,\\
\bSt_i &=  \int_{\partial\fB_i} \overbracket{\br_i\bf} \d S.
\label{forcemoments}
\end{align}

Now taking moments of the flow over the surface of particle $\alpha$, $\bu(\bx\in\partial\fB_\alpha)$, we systematically obtain mobility relationships for the $\alpha$ particle
\begin{align}
\bU_\alpha+\bU_\alpha^s-\bU_\alpha^\infty &=  -\bM^{\alpha\alpha}_{UF}\cdot\bF_\alpha-\sum_{\beta\ne\alpha}\bigg[ \bM^{\alpha\beta}_{UF}\cdot\bF_\beta+\bM^{\alpha\beta}_{UL}\cdot\bL_\beta+\bM^{\alpha\beta}_{US}:\bS_\beta+\ldots\bigg],\\
\bOmega_\alpha+\bOmega_\alpha^s-\bOmega_\alpha^\infty &=  -\bM^{\alpha\alpha}_{\Omega L}\cdot\bL_\alpha-\sum_{\beta\ne\alpha}\bigg[ \bM^{\alpha\beta}_{\Omega F}\cdot\bF_\beta+\bM^{\alpha\beta}_{\Omega L}\cdot\bL_\beta+\bM^{\alpha\beta}_{\Omega S}:\bS_\beta+\ldots\bigg],\\
\bE_\alpha^s-\bE_\alpha^\infty &=  -\bM^{\alpha\alpha}_{ES}\cdot\bS_\alpha-\sum_{\beta\ne\alpha}\bigg[ \bM^{\alpha\beta}_{EF}\cdot\bF_\beta+\bM^{\alpha\beta}_{EL}\cdot\bL_\beta+\bM^{\alpha\beta}_{ES}:\bS_\beta+\ldots\bigg],
\end{align}
where the stresslet for active particles includes a contribution from the double-layer kernel
\begin{align}
\bS_\beta = \bSt_\beta -2\eta \frac{4\pi a_\beta^3}{3}\bE^s_\beta,
\end{align}
while the mobility tensors are identical to those for passive particles
\begin{align}
\bM^{\alpha\alpha}_{UF}&=\frac{1}{6\pi a_\alpha \eta}\bI,\\
\bM^{\alpha\beta}_{UF}&=\frac{1}{4\pi a_\alpha^2}\int_{\partial\fB_\alpha}\d S(\bx)\frac{1}{4\pi a_\beta^2}\int_{\partial\fB_\beta}\bG(\bx,\by)\d S(\by),\\
\bM^{\alpha\beta}_{UL}&=\frac{1}{4\pi a_\alpha^2}\int_{\partial\fB_\alpha}\d S(\bx)\frac{3}{8\pi a_\beta^3}\int_{\partial\fB_\beta}\bG(\bx,\by)\times\bn(\by)\d S(\by),\\
\bM^{\alpha\beta}_{US}&=\frac{1}{4\pi a_\alpha^2}\int_{\partial\fB_\alpha}\d S(\bx)\frac{3}{8\pi a_\beta^3}\int_{\partial\fB_\beta}\bG(\bx,\by)(\bn(\by)+\bn(\by)^\top)\d S(\by),\\
\bM_{\Omega L}^{\alpha\alpha}&=\frac{1}{8\pi\eta a_\alpha^3}\bI,\\
\bM_{\Omega F}^{\alpha\beta}&=\frac{3}{8\pi a_\alpha^3}\int_{\partial\fB_\alpha}\d S(\bx)\frac{1}{4\pi a_\beta^2}\int_{\partial\fB_\beta}\d S(\by)\bn(\bx)\times\bG(\bx,\by),\\
\bM_{\Omega L}^{\alpha\beta}&=\frac{3}{8\pi a_\alpha^3}\int_{\partial\fB_\alpha}\d S(\bx)\frac{3}{8\pi a_\beta^3}\int_{\partial\fB_\beta}\d S(\by)\bn(\bx)\times\bG(\bx,\by)\times\bn(\by),\\
\bM_{\Omega S}^{\alpha\beta}&=\frac{3}{8\pi a_\alpha^3}\int_{\partial\fB_\alpha}\d S(\bx)\frac{3}{8\pi a_\beta^3}\int_{\partial\fB_\beta}\d S(\by)(\bn(\bx)\times\bG(\bx,\by)(\bn(\by)+\bn(\by)^\top),\\
\bM^{\alpha\alpha}_{ES}&=\frac{3}{20\pi\eta a_\alpha^3}\bbI,\\
\bM_{E F}^{\alpha\beta}&=\frac{3}{8\pi a_\alpha^3}\int_{\partial\fB_\alpha}\d S(\bx)\frac{1}{4\pi a_\beta^2}\int_{\partial\fB_\beta}\d S(\by)(\bn(\bx)+\prescript{\top}{}\bn(\bx))\bG(\bx,\by),\\
\bM_{E L}^{\alpha\beta}&=\frac{3}{8\pi a_\alpha^3}\int_{\partial\fB_\alpha}\d S(\bx)\frac{3}{8\pi a_\beta^3}\int_{\partial\fB_\beta}\d S(\by)(\bn(\bx)+\prescript{\top}{}\bn(\bx))\bG(\bx,\by)\times\bn(\by),\\
\bM_{E S}^{\alpha\beta}&=\frac{3}{8\pi a_\alpha^3}\int_{\partial\fB_\alpha}\d S(\bx)\frac{3}{8\pi a_\beta^3}\int_{\partial\fB_\beta}\d S(\by)((\bn(\bx)+\prescript{\top}{}\bn(\bx))\bG(\bx,\by)(\bn(\by)+\bn(\by)^\top),
\end{align}
where $\bbI$ is the fourth-order identity tensor. We give the mobilities here in integral form but it is much more common to see them in the equivalent differential form, which may be found by Taylor expansion about the particle centres \citep{wajnryb13, mizerski14, fiore17}.

For all $N$ particles we write the mobility relationships between velocity moments and force moments in compact form
\begin{align}
\tU+\tU^s-\tU^\infty &= -\tM_{\tU\tF}\cdot\tF-\tM_{\tU\tS}:\tS+\ldots\label{mobilityU}\\
\tE^s-\tE^\infty &= -\tM_{\tE\tF}\cdot\tF-\tM_{\tE\tS}:\tS+\ldots\label{mobilityE}\\
&\vdots\nonumber
\end{align}
In this way we form the grand mobility tensor (typically truncated at the $\tE,\tS$ level shown above in Stokesian Dynamics)
\begin{align}
\fU' = -\fM\cdot\fF.
\end{align}

\section{Dilute approximation}\label{app:dilute}
Dilute approximations are typically used throughout the literature in order to avoid the computational expense of inverting the far-field grand mobility tensor. As shown above in \eqref{mobilityU}, the far-field contribution to the swimming dynamics may be written in terms of mobilities. The stresslet is not prescribed but induced, so solving \eqref{mobilityE} for $\tS$ and substituting into \eqref{mobilityU} yields
\begin{align}
\tU &= \tU^\infty-\tU^s -(\tM_{\tU\tF}-\tM_{\tU\tS}:\tM_{\tE\tS}^{-1}:\tM_{\tE\tF})\cdot\tF+\tM_{\tU\tS}:\tM_{\tE\tS}^{-1}:(\tE^s-\tE^\infty)+\ldots,
\end{align}
and for force-free particles we have simply
\begin{align}
\tU &= \tU^\infty-\tU^s+\tM_{\tU\tS}:\tM_{\tE\tS}^{-1}:(\tE^s-\tE^\infty)+\ldots \ .
\end{align}
To leading order in a dilute approximation only the single particle stresslet terms remain, $\tM_{\tE\tS}^{-1}$ is diagonal, and then we have
\begin{align}
\bU_\alpha &=\bU_\alpha^\infty-\bU_\alpha^s  -\sum_{\beta\ne\alpha}\bM^{\alpha\beta}_{US}:\bS_\beta+\ldots,\\
\bOmega_\alpha &=\bOmega_\alpha^\infty-\bOmega_\alpha^s  -\sum_{\beta\ne\alpha}\bM^{\alpha\beta}_{\Omega S}:\bS_\beta+\ldots,
\end{align}
where
\begin{align}
\bS_\beta=(\bM^{\beta\beta}_{ES})^{-1}:\lb(\bE_\beta^\infty-\bE_\beta^s\rb)&=\frac{20\pi\eta a_\beta^3}{3}\lb(\bE_\beta^\infty-\bE_\beta^s\rb).
\end{align}

\bibliography{swimming}
\end{document}